\documentclass[sigconf]{acmart}
\usepackage{enumitem}
\usepackage[utf8]{inputenc}
\usepackage{mathtools, nccmath}
\usepackage{makecell}
\RequirePackage[T1]{fontenc} \RequirePackage[tt=false, type1=true]{libertine} \RequirePackage[varqu]{zi4} \RequirePackage[libertine]{newtxmath}

\AtBeginDocument{%
  \providecommand\BibTeX{{%
    \normalfont B\kern-0.5em{\scshape i\kern-0.25em b}\kern-0.8em\TeX}}}

\copyrightyear{2023}
\acmYear{2023}
\setcopyright{rightsretained}
\acmConference[CIKM '23]{Proceedings of the 32nd ACM International Conference on Information and Knowledge Management}{October 21--25, 2023}{Birmingham, United Kingdom}
\acmBooktitle{Proceedings of the 32nd ACM International Conference on Information and Knowledge Management (CIKM '23), October 21--25, 2023, Birmingham, United Kingdom}
\acmDOI{10.1145/3583780.3615229}
\acmISBN{979-8-4007-0124-5/23/10}

\makeatletter
\gdef\@copyrightpermission{
  \begin{minipage}{0.3\columnwidth}
   \href{https://creativecommons.org/licenses/by/4.0/}{
        \includegraphics[width=0.90\textwidth]{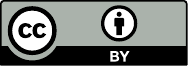}
    }
  \end{minipage}\hfill
  \begin{minipage}{0.7\columnwidth}
   \href{https://creativecommons.org/licenses/by/4.0/}{This work is licensed under a Creative Commons Attribution International 4.0 License.}
  \end{minipage}
  \vspace{5pt}
}
\makeatother

\begin{document}
\setlength{\abovedisplayskip}{3pt}
\setlength{\belowdisplayskip}{2pt}


\title{Efficient Multi-Task Learning via Generalist Recommender}


\author{Luyang Wang}
\email{luyang.wang@verizon.com}
\affiliation{
  \institution{Verizon}
  \city{Basking Ridge, NJ}
  \country{USA}}

\author{Cangcheng Tang}
\email{cangcheng.tang@verizon.com}
\affiliation{
  \institution{Verizon}
  \city{Boston, MA}
  \country{USA}}

\author{Chongyang Zhang}
\email{chongyang.zhang@intel.com}
\affiliation{
  \institution{Intel Corporation}
  \city{Santa Clara, CA}
  \country{USA}}

\author{Jun Ruan}
\email{jun.ruan@verizon.com}
\affiliation{
  \institution{Verizon}
  \city{Alpharetta, GA}
  \country{USA}}

\author{Kai Huang}
\email{kai.k.huang@intel.com}
\affiliation{
  \institution{Intel Corporation}
  \city{Santa Clara, CA}
  \country{USA}}

\author{Jason Dai}
\email{jason.dai@intel.com}
\affiliation{
  \institution{Intel Corporation}
  \city{Santa Clara, CA}
  \country{USA}}

\renewcommand{\shortauthors}{Luyang Wang et al.}


\begin{abstract}
Multi-task learning (MTL) is a common machine learning technique that allows the model to share information across different tasks and improve the accuracy of recommendations for all of them. Many existing MTL implementations suffer from scalability issues as the training and inference performance can degrade with the increasing number of tasks, which can limit production use case scenarios for MTL-based recommender systems. Inspired by the recent advances of large language models, we developed an end-to-end efficient and scalable Generalist Recommender (GRec). GRec takes comprehensive data signals by utilizing NLP heads, parallel Transformers, as well as a wide and deep structure to process multi-modal inputs. These inputs are then combined and fed through a newly proposed task-sentence level routing mechanism to scale the model capabilities on multiple tasks without compromising performance. Offline evaluations and online experiments show that GRec significantly outperforms our previous recommender solutions. GRec has been successfully deployed on one of the largest telecom websites and apps, effectively managing high volumes of online traffic every day.
\end{abstract}

\ccsdesc[500]{Information systems~Recommender systems}
\ccsdesc[500]{Computing methodologies~Neural networks}

\keywords{Multi-Task Learning; Recommender Systems; Real-World Application}

\maketitle
\section{Introduction}
When developing a recommender system (RS), the recommendation task can be viewed as a next-item prediction problem, where the goal is to optimize various performance metrics given by a set of user behavior and contextual information. There are a wide variety of performance metrics that can be optimized by RS, such as click through rate (CTR), add to cart rate (ATC), conversion rate (CVR), etc. Even the same performance metrics optimization can greatly differ depending on the context and specific use case scenarios. For example, predicting which smartphone a user would purchase next can be defined as a CVR task. Additionally, it can be subdivided into two distinct scenarios: acquiring a smartphone through a device trade-in flow or obtaining a smartphone by adding a new line to the wireless account. 

The recent development of MTL has demonstrated promising performance that allows one model to optimize across multiple tasks \cite{MMoE}. However, scaling challenges arise in many existing MTL architectures as training and inference speeds degrade when the number of tasks increases. Inference performance is particularly important in real-world recommender systems, and limitations in this aspect can restrict the application of existing MTL-based recommender systems when scaling to multiple tasks.

As one of the largest telecommunication companies in the world, previously we have developed many single-task recommenders for individual use cases to optimize different performance metrics. This approach has led to several challenges. First, models working in silos may fail to consider the interconnection among various use cases, resulting in a narrow model vision and potential recommendation bias. Second, training data is sometimes sparse for certain tasks, such as CVR-related tasks. Insufficient training data presents challenges for models with large numbers of parameters to optimize. Third, maintaining multiple single-task recommenders can increase the complexity of ML operations. 

To overcome the above challenges, we proposed Generalist Recommender (GRec) that can handle multiple recommender tasks simultaneously. Taking inspiration from the recent development of LLMs \cite{GLaM}, we applied the sparse mixture-of-experts \cite{sparsemoe} (sparse MoE) architecture and proposed a new task-sentence routing strategy, allowing our model to expand its capacity for cross-task generalization while maintaining promising inference performance. 

Our primary contributions can be summarized as follows:
\begin{itemize}
\item We proposed a novel recommender GRec which is able to generalize across a variety of recommendation tasks.
\item We applied sparse MoE architecture and proposed a new task-sentence routing strategy to efficiently scale up GRec in real-world applications.
\item We conducted experiments on a real-world large-scale digital recommender system, demonstrating that GRec delivers significant performance and efficiency gains.
\end{itemize}

\section{Related Work}
\textbf{Multi-Task Learning for Recommendation:} To address the challenge of multiple objectives in personalized recommendation scenarios, such as clicking, adding to cart, and purchasing, many multi-task models \cite{OMoE, MMoE, SNR, MetaHybridExperts} have been proposed that can jointly learn from several tasks and improve the accuracy of recommendations. However, existing multi-task approaches usually serve for scenarios with a small number of tasks\cite{MetaHybridExperts} and are not suitable for large-scale online recommendation \cite{NextVideo, MUVE, MulViewDL, ExpRec}. Due to the large number of task-specific parameters of existing models, the input modalities of these multi-task methods cannot be scaled up \cite{NextVideo}. 

\textbf{Large-Scale Applications of Multi-Task Methods:} With the development of distributed machine learning systems \cite{dean2012large}, large-scale models represented by LLMs are increasingly used in real-life applications, such as GPT-3 \cite{GPT-3}, PaLM \cite{palm}, LLaMa \cite{llama}. In addition to recommendation systems, large-scale models with multi-task learning capabilities have also been applied to traditional machine learning tasks \cite{taskmoe, multilinearRela, sparsemoe}. For example, Kudugunta et al. \cite{taskmoe} proposed Task-MoE for multilingual machine translation tasks, which can extract smaller, ready-to-deploy sub-networks from large sparse models. With task-level routing, Task-MoE selects experts by task boundaries as opposed to making input-level decisions, significantly improving efficiency and scalability. Inspired by LLMs and Task-MoE, our GRec can select the required expert combination based on task and has the ability to deal with tasks of different modalities. Through extensive experiments on real-world large-scale recommendation systems, we demonstrate that our GRec is indeed scalable and has impressive performance.

\section{MODEL ARCHITECTURE}

\begin{figure}[t]
  \centering
   \includegraphics[width=0.9\linewidth]{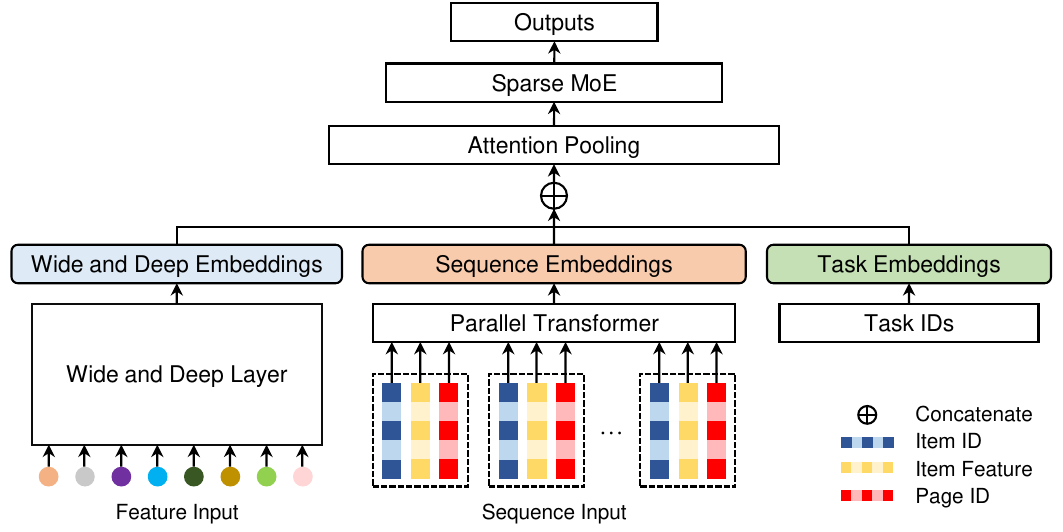}
   \caption{The framework of our GRec model.
   }
   \setlength{\parindent}{0pt}
   \label{fig:model}
   \vspace{-10pt}
\end{figure}

The overall architecture of our GRec is illustrated in Fig. \ref{fig:model}, while this section provides an elaborate description of the pivotal components of GRec.

\begin{figure}[t]
  \centering
   \includegraphics[width=1\linewidth]{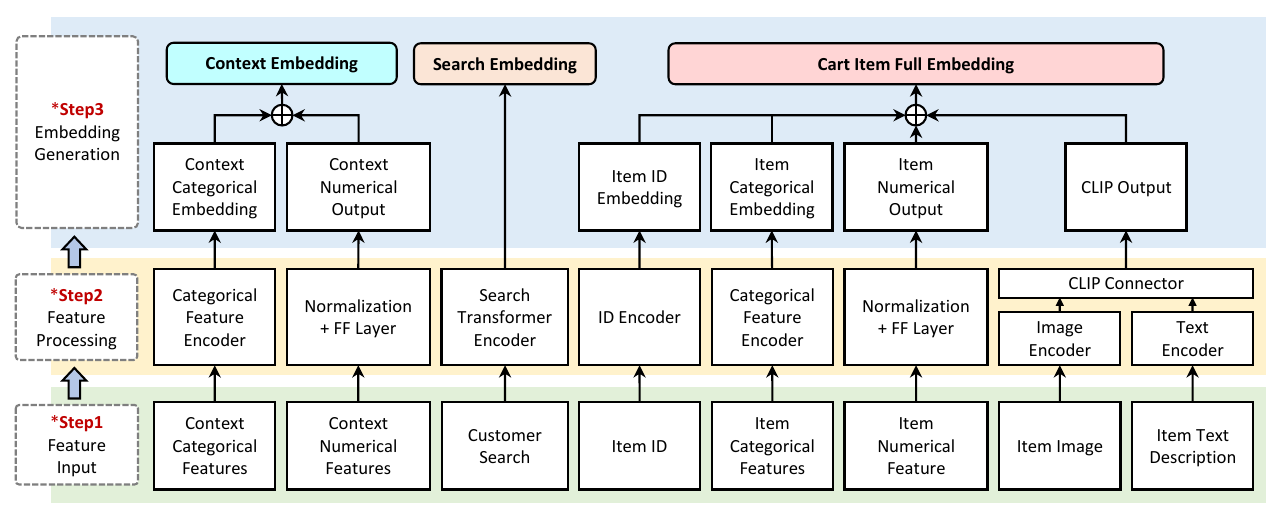}
   \caption{The architecture of wide and deep layer in GRec.}
   \label{fig:wideandeep}
   \vspace{-10pt}
\end{figure}

\subsection{Wide and Deep Layer}
\label{wdl}
A wide and deep structure \cite{wideandeep} is used to process inputs of multi-modalities, including categorical and numerical data, texts, and images, as shown in Fig. \ref{fig:wideandeep}. Customer categorical features are encoded into embedding vectors and passed into the deep tower. Numerical features are processed using a feedforward layer to ensure the output dimension matches that of the deep tower. GRec passes search terms into a pre-trained transformer encoder to capture customer intent. On the item side, item meta features are utilized alongside item ID embedding to enrich the item encoding layer. Item categorical features are converted into embeddings (item deep component), while numerical features are normalized and passed into a feedforward layer (item wide component). GRec leveraged embedding input extracted from a pre-trained CLIP model \cite{CLIP} to encode item image and text description. Item ID embeddings, item wide and deep components, and CLIP model outputs are then concatenated to form a full item embedding vector.

\subsection{Parallel Transformer Layer}
Sequence data includes past devices  that customers viewed and past pages they landed on our websites. GRec uses parallel attention and feedforward with residual connection to encode those sequences. This is an adaption from the Pathways Language Model \cite{palm}, which contains below highlights:

\textbf{Multi-query single-key-value attention.} Traditionally, the scaled dot-product attention \cite{transformer} can be formulated as:

\begin{equation}
\label{eq-1}
\mathrm{Attention} (Q,K,V)=\mathrm{softmax} (\frac{QK^{T}}{\sqrt{d}})V
\end{equation}

\noindent where $Q$ represents the queries, $K$ the keys, $V$ the values and $d$ the dimension of latent vectors. Following \cite{transformer}, we use multi-head attention:

\begin{equation}
\begin{aligned}
\operatorname{MultiHead}(Q, K, V) & =\operatorname{Concat}\left(\operatorname{head}_1, \ldots, \operatorname{head}_{\mathrm{h}}\right) W^O, \\
\operatorname{head}_{\mathrm{i}} & =\operatorname{Attention}\left(Q W_i^Q, K W_i^K, V W_i^V\right)
\end{aligned}
\end{equation}

\noindent where the projection matrices $W^Q$, $W^K$, $W^V\in R^{d\times d}$, and $h$ is the number of heads. For a standard Transformer with h attention heads, the shape of the $Q$, $K$, and $V$ tensors is $[h, d]$, where $d$ is the attention head size. 
Different from the aforementioned multi-head attention mechanism, for different queries, we are using the same key and value. That is,  the key/value projections are shared for each head, i.e. $K$ and $V$ are projected to $[1, d]$, but $Q$ is still projected to shape $[h, d]$. This improves efficiency by reducing attention block size and computation complexity.

\textbf{Parallel Layer.} Following \cite{palm}, we use the parallel layer in each transformer block, which parallelizes the attention and feedforward layers. The formula of traditional transformer can be written as:
\begin{equation}
y = x+\mathrm{MLP} (\mathrm{LayerNorm} (x+\mathrm{Attention} (\mathrm{LayerNorm} (x)))
\end{equation}

\noindent whereas the formula of parallel transformer can be written as:

\begin{equation}
y = x+\mathrm{MLP} (\mathrm{LayerNorm}(x))+\mathrm{Attention} (\mathrm{LayerNorm} (x))
\end{equation}

This architecture is able to reduce model complexity and increase attention speed, especially at large scale \cite{palm}. 

\subsection{Task-Sentence MoE Layer} 

\begin{figure*}[t]
  \centering
   \includegraphics[width=0.7 \linewidth]{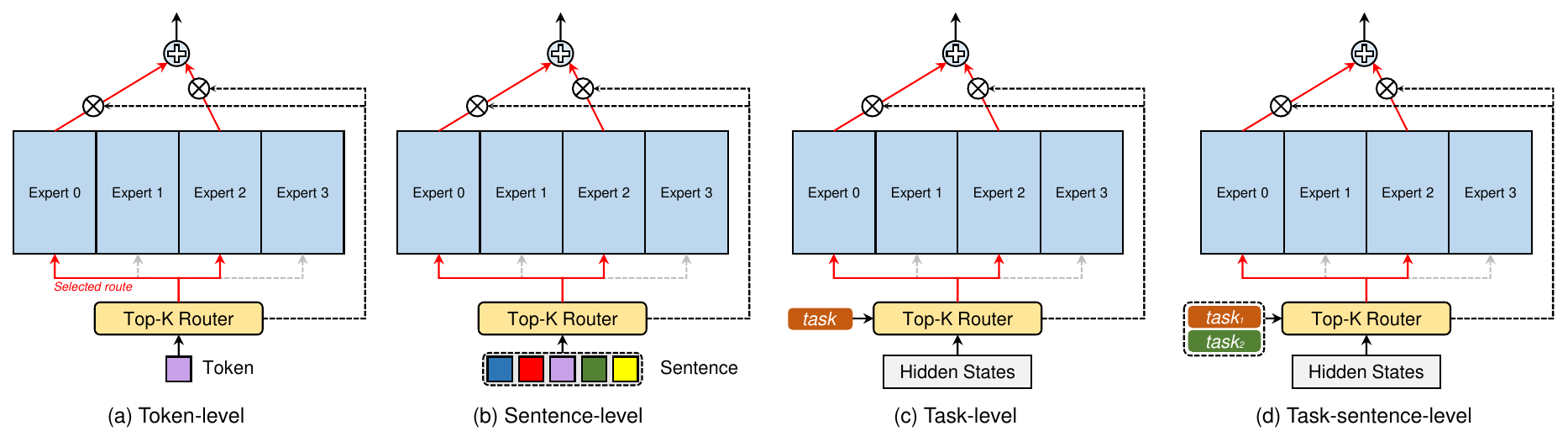}
   \caption{Differences between token-level, sentence-level, task-level and task-sentence-level sparse MoEs.
   }
   \label{fig:moes}
   \vspace{-10pt}
\end{figure*}

To enhance GRec’s capability to generalize across multiple recommendation task categories, we scale up GRec model parameters using the sparse MoE \cite{sparsemoe} structure. This structure is capable of activating a subset of expert layers depending on task categories,  which allows multiple tasks to be combined and trained in one model. 

For the MoE layer, the gating function (also referred to as routing strategy) is critical, which indicates the weights of each expert in processing incoming tokens \cite{gshard}. In multilingual translation, some common routing strategies for MoE include (i) token-level routing, (ii) sentence-level routing and (iii) task-level routing \cite{taskmoe}, as detailed below.

\textbf{Token-Level Routing.} In token-level routing, each token is routed independently, as shown in Fig. \ref{fig:moes} (a):

\begin{equation}
\mathcal{G}_{s,E}=\mathrm{GATE} (x_s)
\end{equation}



\noindent where $x_s$ is the input token to the MoE layer. Vector $\mathcal{G}_{s,E}$ is computed by a gating network (also referred as router). We use this vector to select a subset of experts to route the token. 

\textbf{Sentence-Level Routing.} As shown in Fig. \ref{fig:moes} (b), all tokens from a sentence are routed to the same experts, and the gating vector is calculated by concatenating all token representations in a given sentence:

\begin{equation}
\mathcal{G}_{s,E}=\mathrm{GATE} (\frac{1}{S} \sum_{s=1}^S x_s)
\end{equation}

\textbf{Task-Level Routing.} Experts are selected by task boundaries. In multilingual translation, task boundaries can be defined by the target language or the language pair, the structure of task-level routing is shown in Fig. \ref{fig:moes} (c). Task-level routing is formulated as follows: 

\begin{equation}
\mathcal{G}_{s,E}=\mathrm{GATE} (\mathrm{task_{id}})
\end{equation}

\noindent where $\mathrm{task_{id}}$ is a manually set input that represents the current task to be processed.


Different from multilingual translation, in the field of recommendation, routing tasks can have multiple types, in our case, it has two types: flow and use cases. Customer digital interactions typically follow three flows: adding a line to an existing account (AAL), upgrading current devices (EUP), and prospect customers’ acquisition of new services (NSE). Use cases here refer to the business goal that is being targeted, such as CTR and CVR. Pairing these task types will create too many different tasks, and these splits often imbalance the dataset and lead to unstable models. Referring to the routing strategies in multilingual translation, in GRec, we are introducing a new routing strategy: Task-Sentence level routing, which combines multiple task tokens into a sentence and then performs expert routing. In our case, we combine our main task (e.g., CTR, CVR) with the auxiliary task (e.g., EUP, AAL) as a task sentence (e.g., AAL+CVR, EUP+CTR) to feed into routing. With this approach, embeddings from different types of tasks are considered without creating too many task type pairs. We define each task based on user ordering flow (device upgrade, add a line, new customer, etc.) and targeted outcome (CTR, CTCVR, CVR, etc) pair, as shown in Fig. \ref{fig:moes} (d). The formula of the \textbf{task-sentence} routing strategy is as follows:

\begin{equation}
\mathcal{G}_{s,E}=\mathrm{GATE} (\frac{1}{S} \sum_{s=1}^S \mathrm{task_{id}}_s)
\end{equation}

\noindent After comparing the performance trade-offs between different routing strategies (as shown in Sec. 4), we adopt task-sentence level routing strategies in the GRec implementation. 

\section{EXPERIMENTS} \label{experiments}
In this section, we perform experiments to evaluate the performance of our proposed framework and compare routing strategies on speed and average precision. Experiments are conducted on public datasets and our internal transaction data. On the public dataset, we conduct separate experiments on sparse MoE in GRec to validate its scalability. On our internal transaction data, we conducted offline comparison on various routing strategies and use MMoE\cite{MMoE} as baseline. We also conducted online A/B testing of GRec over previous state-of-art method and achieved impressive performance in real-world large-scale recommender applications.

\subsection{Offline evaluation on AliExpress dataset}
\textbf{Dataset:} For the public dataset experiment, we conducted on AliExpress dataset \cite{aliexpress}, which is gathered from real-world traffic logs of research system. This dataset is collected from 5 countries: Russia (RU), Spain (ES), French (FR), Netherlands (NL), and America (US), which can be seen as 5 independent multi-task sub-datasets. 

\textbf{Settings:} On public dataset, the hyper-parameters shared by all models are set to the same values, the source and hyper-parameter settings are all listed in \cite{BigDL} to facilitate the reproduction of our experiments. Due to the large amount of data in the RU sub-dataset, we mainly conducted experiments on four other sub-datasets, taking the AUC \cite{AUc} score as the metric. Note a slightly higher AUC at 0.1\%-level is regarded as significant for the CTR task \cite{eee}. 

\begin{figure}[t]
  \centering
   \includegraphics[width=0.9\linewidth]{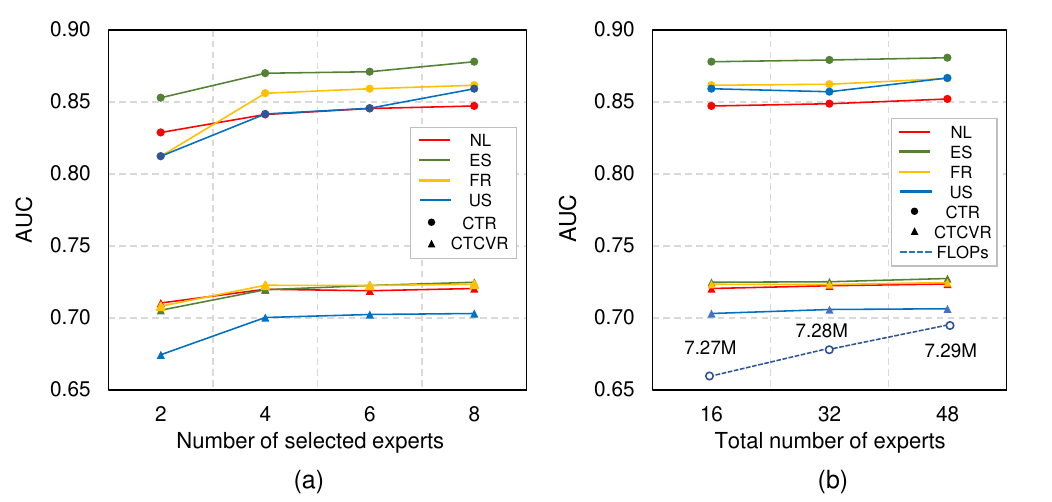}
   \caption{The trend of AUC changes when the number of experts selected and the total number of experts are different.}
   \label{fig:experiment}
   \vspace{-10pt}
\end{figure}

As shown in Fig. \ref{fig:experiment}, as the number of selected experts (the $k$ of top-$k$ router) increases and the total number of experts changes, the performance of the model will be improved. This means that we can not only improve performance by changing top-$k$, but also by increasing the total number of experts. At the same time, when increasing the total number of experts, the FLOPs (floating point operations per second) of the model shows slight changes, further verifying the scalability of sparse MoE in recommendation tasks. This is also the reason why we apply sparse MoE in GRec.

\begin{table}[hbt]
\small
\begin{center}
\centering
\setlength{\tabcolsep}{2.pt}
\caption{Results on internal offline transaction dataset. 
}
\label{table:offline_internal_results}
\begin{tabular}{c|c|c|c|c|c|c|c}
\hline
System & \makecell{Expert / \\ Top $k$} & \makecell{Routing \\ Granularity} & \makecell{FLOPs \\ (Routing)} & \multicolumn{4}{c}{Average Precision} \\
\cline{5-8}
 &  &  &  & CTR & CTCVR & ATC & CVR \\
\hline
MMoE & 4/4 & Sentence & 2.1M & 64.15 & 79.02 & 55.73 & 84.17\\
\hline
GRec & 8/4 & Token & 4.1M & \textbf{64.53} & \textbf{80.72} & 56.23 & 84.63 \\
\cline{2-8}
 & 8/4 & Sentence & 4.1M & 64.12 & 78.95 & 55.12 & 83.67 \\
\cline{2-8}
 & 8/4 & Task & 1.0M & 64.26 & 79.24 & 55.21 & 84.24 \\
\cline{2-8}
 & 8/4 & Task-Sentence & 2.1M & 64.51 & 80.71 & \textbf{56.85} & \textbf{84.94} \\
 \hline
\end{tabular}
\end{center}
\vspace{-30pt}
\end{table}

\subsection{Offline evaluation of internal transaction dataset}
\textbf{Dataset:} For our internal experiment, GRec is trained on 3 months of customer transaction data, and test data is sampled on transactions that happened one week after the last day of training data. The two types of tasks are \textbf{flows}: EUP, AAL, NSE, and \textbf{use cases}: CTR, CTCVR, ATC, CVR. Training data are upsampled by tasks to ensure data balance. 

\textbf{Settings:} Different versions of sparse MoE models are tested, all versions pick top-4 routing on 8 experts, with expert capacity of 2 tokens and 2048 as the batch size. We evaluate model performance based on FLOPs and average precision for the given use cases. We analyze the impact of different sparse MoE routing strategies on model performance, comparing them not only against the baseline but also among themselves.

\textbf{Baseline:} MMoE is used as the baseline model, and it is setup with 4 experts to match the top-4 routing in sparse MoE. 

The results are presented in Table~\ref{table:offline_internal_results}. As the baseline, the MMoE strategy demonstrated lower performance in average precision (AP) and similar FLOPs consumption compared to GRec task-sentence level routing. When comparing MMoE with sparse-MoE, even though routing FLOPs are similar, training and inference time grow linearly on the expert number, as MMoE has to use all its experts instead of choosing top-$k$ flexibly. GRec task-sentence routing showed comparable performance to token-level routing, while exhibiting a notable 50\% improvement in FLOPs. It effectively balanced model performance and efficiencies.

\vspace{-0.15cm}
\begin{table}[hbt]
\small
\centering
\setlength{\tabcolsep}{2.pt}
\caption{Results on internal online A/B test. 
}
\label{table:online_ABtest_results}
\begin{tabular}{l|l|c}
\hline
\hfil Task  & Baseline Recommender & \makecell{GRec Improvement} \\
\hline
\makecell{Home Page \\ (AAL CTCVR)}  &  \makecell{Add a Line \\ Device model} & +8.5\%  \\
\hline
\makecell{Home Page \\ (EUP CTCVR)}  &  \makecell{Upgrade \\ Device model} & +2.1\%  \\
\hline
\makecell{Accessory \\ Interstitial Page \\ (AAL CVR)}  &  \makecell{Add a Line \\ Accessory model} & +20.3\%  \\
\hline
\makecell{Accessory \\ Interstitial Page \\ (EUP CVR)}  &  \makecell{Upgrade \\ Accessory model} & +12.0\%  \\
 \hline
\end{tabular}
\end{table}

\subsection{ Online A/B Testing evaluation}
For online measurement, we evaluated GRec in four different sets of use cases against previous state-of-art single-task recommender models. 
\begin{itemize}
\item The first and second use cases are on the home page, where the recommendation task is defined as to improve the device click-through conversion rate (CTCVR) with different flows of upgrade (EUP) and add a line (AAL).
\item The third and fourth use cases are on the interstitial page, where the recommendation task is defined as to improve accessory conversion with different flows of upgrade (EUP) and add a line (AAL).
\end{itemize}

The single-task recommender is trained on the specific task only. Each use case was evaluated separately using the same amount of live traffic.

Table~\ref{table:online_ABtest_results} shows the results of online A/B testing. We observed significant improvements across all tasks during the online A/B testing period, and all reached statistical significance. It’s worth noting that GRec achieved an even higher improvement on CVR-related tasks over single task recommender, where conversion-related tasks are relatively sparse and we see GRec benefited a lot from shared parameter learning from other tasks with more abundant training data such as CTR and ATC.

\section{Conclusion and Future Works}

In this paper, we purpose a new recommender model, GRec, which applies the sparse-MoE structure and utilizes our novel task-sentence routing strategy. Moreover, our model is designed to take inputs of multi-modalities, which facilitates generalizing tasks. In multiple production use cases, GRec has shown significantly improved performance over the baseline in both offline and online A/B testing settings. We demonstrate success deploying GRec in real-world large-scale recommender systems. 

\vspace{0.25cm}
\bibliographystyle{siamplain}
\balance
\bibliography{reference}

\end{document}